# Measuring Consumer Perceived Warm-Glow for Technology Adoption Modeling

Antonios Saravanos

New York University, saravanos@nyu.edu

Dongnanzi Zheng

New York University, dz40@nyu.edu

Stavros Zervoudakis

New York University, zervoudakis@nyu.edu

In this paper, we adapt and validate two constructs—perceived extrinsic warm-glow (PEWG) and perceived intrinsic warm-glow (PIWG)—to measure the two dimensions of consumer perceived warm-glow (i.e., extrinsic and intrinsic) for use with the practice of technology adoption modeling. Taking an experimental approach, participants were exposed to one of four vignettes designed to simulate either the absence or the presence of warm-glow (specifically, extrinsic warm-glow, intrinsic warm-glow, and concurrently extrinsic and intrinsic warm-glow). The results revealed that both constructs measured their respective forms of warm-glow with two caveats. Firstly, singularly trying to evoke extrinsic warm-glow led to only a slight increase in consumer perception of extrinsic warm-glow. We attributed this finding to individuals not being attracted to technology products that overtly target and seek to satisfy their vanity, instead preferring technology that does so in a more subtle way. The second is that singularly trying to evoke intrinsic warm-glow also resulted in the manifestation of extrinsic warm-glow. Thus, warm-glow appears as a blend of extrinsic and intrinsic dimensions. This finding serves to reinforce what has already been reported in existing literature regarding warm-glow and the idea of impure altruism.

## 1 INTRODUCTION AND BACKGROUND

The practice of technology adoption modeling is described as being well-established and mature [1], finding its origins in the work of Davis [2] and his original 'Technology Acceptance Model', known colloquially as TAM. Over time, two 'standard' approaches to technology adoption modeling have emerged. The first endeavors to develop the original TAM model for which there have been three further evolutions (i.e., TAM1 [3], TAM2 [4], and TAM3 [5]), with the most current being, as one would expect, TAM3. The second approach is Venkatesh et al.'s [6] Unified Theory of Acceptance and Use of Technology (UTAUT), the latest version of which is UTAUT2 [7]. Both the TAM3 and UTUAT2 models are, by their nature, flexible, and they have been extended and adapted for a wide variety of cases in order to meet the needs of different technologies and the contexts in which they are used [8–10]. Accordingly, the aim of this work is to adapt and validate constructs to measure consumer perceived warm-glow for use with the practice of modeling consumer adoption, specifically with regard to technology that evokes such a feeling (i.e., the feeling of warm-glow in consumers). For the purposes of this paper, we refer to this as 'good tech'.

The term "warm-glow" was coined by Andreoni [11], who observed that individuals experienced a feeling of "warm-glow" after they had donated to those less fortunate, and thus, "done their bit" for society. Andreoni [11] recognized that consumers who tend to seek out warm-glow could have altruistic as well as non-altruistic motivations for their behavior, explaining that "people have a taste for giving: perhaps they receive status or acclaim". Pure altruistic behavior can be described as "a willingness to benefit others, even at one's own expense" [12]. Non-altruistic behavior, or rather impure altruistic behavior, can be described as acting in a way "that benefits others in order to feel pride in acting altruistically and to avoid the shame of acting selfishly" [12]. These two dimensions of warm-glow are classified as either extrinsic warm-glow (EWG) to reflect a non-altruistic motivation [13], or intrinsic warm-glow (IWG) to reflect an altruistic motivation [14]. Respectively, we offer two constructs to measure these two dimensions of warm-glow: 'Perceived Extrinsic Warm-Glow' (PEWG) and 'Perceived Intrinsic Warm-Glow' (PIWG).

In the remainder of this paper we describe these constructs in greater detail, along with the process that was taken to validate them. We conclude with a discussion of the findings, our contribution to the field and implications, and ways in which the work can be further developed.

## 2 MATERIALS AND METHODS

In this section we first explain the development of the PEWG and PIWG constructs. Next, we describe the experiment that was carried out in order to validate the constructs. After this, we outline the data collection process employed and present the structure of our sample. Lastly, we conclude our work by outlining the approach that was utilized for the analysis of the data collected.

### 2.1 Development of Constructs

As outlined in the introduction, warm-glow is comprised of two dimensions: extrinsic and intrinsic. To measure consumer perception of extrinsic warm-glow we formed the PEWG construct, using questions based on the three items that were provided as part of the construct of "Social Value, Status", developed by Han et al. [15]. To form the PIWG construct, we adapted three questions originating from Müller et al.'s [16] factor of (intrinsic) warm-glow. All questions were rated through a 7-point Likert scale ranging from 'strongly disagree' to 'strongly agree', thereby aligning with the scales used in the TAM3 and UTATU2 instruments. The questions used for both the PIWG and PEWG factors can be seen in Table 1.

### 2.2 Experimental Design

To evaluate the proposed theoretical framework a vignette-based experimental design was taken, inspired by the approach described by Ongena et al. [17]. Internet search technology was selected for the experiment because it is familiar to almost all users and could be easily described in a way that evokes IWG, EWG, or both. Each experiment began by asking participants to confirm that they were willing to participate in the study. Those that consented were permitted to continue, and were asked questions to record their gender, age, income, schooling, race, and prior experience with technology – in particular, their use of internet search technology. Participants were then assigned to one of four vignettes/conditions:

1. CONTROL – the vignette describes an internet search technology as follows: "Consider a free web-based search engine, which is able to mimic your favorite search engine in terms of functionality, and is produced by a new organization". It then states that it "is identical to the web-based Google search engine in terms of functionality", but that "it is not produced by Google but rather by a new organization".

2. EWG – includes language to evoke extrinsic warm-glow in participants: "Your use of the tool results in posts to any of your social media platforms announcing that you are supporting any charity of your choice".
3. IWG – includes language to evoke intrinsic warm-glow in participants: "Your use of the tool results in an anonymous donation to any charity of your choice".
4. BOTH – this last condition presents a vignette that includes the material from both the EWG and IWG conditions to evoke both forms of warm-glow.

Table 1: Items for the Factors of Perceived Extrinsic and Intrinsic Warm-Glow.

| Factor | Item | Questions |
|---|---|---|
| Perceived Extrinsic Warm-Glow (PEWG), based on Han et al.'s [15] factor of Social Value, Status | PEWG1 | Using this search engine fits the impression that I want to give to others that I am a good person whose actions have a positive impact on society. |
| | PEWG2 | I am eager for my friends/acquaintances to learn about my use of this search engine and how I have a positive impact on society through its use. |
| | PEWG3 | I can express a more distinctive personal image through the use of this search engine by demonstrating to others that I am a good person who positively impacts society. |
| Perceived Intrinsic Warm-Glow (PIWG), adapted from Müller et al. [16] | PIWG1 | If I use this search engine, I will feel good because I do not only spend money for myself but also for other people. |
| | PIWG2 | I feel comfortable if I donate for a good cause by using this search engine. |
| | PIWG3 | I am pleased that I do not only get a service by using this search engine, but that I also do a good deed at the same time. |

Subsequently, participants completed a questionnaire sharing their perceptions of the technology that had been presented to them through their respective vignette. To address the possibility that participants might not be fully invested in taking our questionnaire, we selected four questions from Abbey and Meloy [18], which we modified to gauge participant attention. The attention check questions were also rated through a 7-point Likert scale ranging from 'strongly disagree' to 'strongly agree'. The vignettes and questionnaire were distributed through the use of the Qualtrics online survey platform.

## 2.3 Sample and Data Collection

Participants were recruited through the Amazon Mechanical Turk crowdsourcing platform, which has become quite popular for such studies [19,20]. We collected a total of 369 responses, with all participants coming from the United States. Of those, 29 submissions were removed from the final dataset, because they failed to pass the attention checks, leaving 340 remaining responses. The gender distribution of the sample was 173 (50.88%) female, and 167 (49.12%) male, and the majority of participants were in the 31–55 age bracket (77.06%) with some form of either undergraduate (73.53%) or postgraduate (10.59%) training. The breakdown can be seen in greater detail in Table 2. Almost all participants appeared to be frequent users of search engine technology, with 96.18% indicating that they use such technology on a daily basis. The majority of participants (89.12%) indicated that their favorite search engine was Google; this was followed by the DuckDuckGo search engine (6.18%), and then Bing (3.82%).

Table 2: Demographics Profile for the Sample.

| Variable | Item | N | % |
|---|---|---|---|
| Gender | Female | 173 | 50.88% |
| | Male | 167 | 49.12% |
| | 18-25 | 9 | 2.65% |
| Age | 26-30 | 41 | 12.06% |
| | 31-55 | 262 | 77.06% |
| | 56 or older | 28 | 8.24% |
| | Less than $19,999 | 37 | 10.88% |
| | $20,000 to $49,000 | 125 | 36.76% |
| Income | $50,000 to $79,999 | 101 | 29.71% |
| | $80,000 to $149,999 | 66 | 19.41% |
| | $150,000 or more | 11 | 3.24% |
| | Less than high-school degree | 1 | 0.29% |
| | High-school diploma or equivalent | 53 | 15.59% |
| Schooling | Some college but no degree | 72 | 21.18% |
| | Associate degree in college | 48 | 14.12% |
| | Bachelor's degree in college | 130 | 38.24% |
| | Graduate degree | 36 | 10.59% |

**2.4 Analysis**

To inform our analysis we looked to the guidelines proposed by MacKenzie et al. [21] on construct measurement and validating techniques within the field of management information systems. They write, "after a construct has been conceptually defined and tentative measures have been developed, one of the next steps is to test whether the measures behave as one would expect them to if they were valid" [21]. The data analysis process we employed to evaluate each of the constructs relied on two stages. In the first stage a measurement model was applied to examine the relationship between the manifest variables, which are directly measured through the survey questions, and their corresponding latent variables, which are inferred by those questions. This enabled us to ascertain whether the manifest variables were useful in effectively measuring the latent variables. To accomplish this the measures of convergent validity, construct reliability, and discriminant validity were assessed based on a confirmatory factor analysis. Factor loadings and average variance extracted (AVE) were used to test convergent validity, which is a measure that reveals the extent to which items are related between reality and theory [22]. Manifest variables with values greater than 0.7 for both measures suggest appropriate convergent validity. To evaluate construct reliability, composite reliability (CR) and Cronbach's Alpha were used. Values greater than 0.7 for both indicate good construct reliability. To measure discriminant validity, we used the Fornell-Larcker Criterion and Cross Loadings. Fornell and Larcker [23] advise that the correlations between each construct should be lower than the square root of the Average Variance Extracted, while Chin [24] prescribes that each cross-loading should be lower than that of all of the indicator's loading.

In the second stage, concurrent validity and sensitivity were examined to ascertain the effects of exposure of PEWG and PIWG to EWG and IWG. Sullivan and Artino [25] highlight the work of Norman [26] writing that "he provides compelling evidence, with actual examples using real and simulated data, that parametric tests not only can be used with ordinal data, such as data from Likert scales". Moreover, they continue by drawing attention to the fact that Norman [26] states "that parametric tests are generally more robust than nonparametric tests", noting that "parametric tests tend to give 'the right answer' even when statistical assumptions—such as a normal distribution of data—are violated, even to an extreme degree" [25]. They conclude that "parametric tests are sufficiently robust to yield largely unbiased answers that are acceptably close to 'the truth' when analyzing Likert scale responses" [25]. To test concurrent validity, the approach of Drisko and Grady [27] was followed, using a Pearson product-moment correlation to determine whether exposure to simulated EWG and IWG in the form of vignettes impacted the rated values of PEWG and PIWG. In order to ascertain the sensitivity of the PEWG and PIWG ratings when exposed to intrinsic and/or extrinsic warm-glow, the approach prescribed by Rosengren et al. [28] was applied. Specifically, we used a multivariate analysis of variance (hereafter referred to as MANOVA) to examine whether the vignette had a statistically significant effect on how participants responded to the questions related to PEWG and PIWG. Should the MANOVA reveal statistical significance for any of the independent variables the next step will be to complete a univariate Analysis of Variance (ANOVA hereafter) with respect to the dependent variables (PEWG and PIWG).

## 3 RESULTS

In this section we outline the results that were obtained.

### 3.1 Measurement Model

To evaluate our measurement model, we assessed the measures of convergent validity, construct reliability, and discriminant validity.

*3.1.1 Convergent Validity*

With respect to convergent validity, we relied on the factor loadings and the average variance extracted (AVE). None of the manifest variables had values lower than 0.7 which satisfied the criteria prescribed by Chin [29]. The items were also statistically significant ($p < 0.05$, t-statistics were obtained from bootstrapping with 7000 subsamples), reflecting that they possessed appropriate convergent validity (see Tables 3 and 4).

Table 3: Factor Loadings of the Reflective Constructs.

| Item  | Loading | t-statistic |
|-------|---------|-------------|
| PEWG1 | 0.948   | 150.281*    |
| PEWG2 | 0.913   | 75.205*     |
| PEWG3 | 0.927   | 85.682*     |
| PIWG1 | 0.930   | 76.610*     |
| PIWG2 | 0.923   | 76.940*     |
| PIWG3 | 0.941   | 77.362*     |

*$p < 0.01$.

Table 4: Average Variance Extracted, Composite Reliability and Alpha Values of the Reflective Constructs.

| Factor | Number of Items | AVE   | CR    | Cronbach's Alpha |
|--------|-----------------|-------|-------|------------------|
| PEWG   | 3               | 0.864 | 0.950 | 0.921            |
| PIWG   | 3               | 0.868 | 0.952 | 0.924            |

*3.1.2 Construct Reliability*

Construct reliability was evaluated through the measures of composite reliability (CR) and Cronbach's Alpha. For both measures, we found values greater than 0.7 with respect to all items, indicating good construct reliability (see Table 4).

*3.1.3 Discriminant Validity*

The Fornell-Larcker criterion revealed that the correlations between each construct were lower than the square root of the AVE (see Table 5). Concerning the cross-loadings, each cross-loading was lower than all of the indicator's loadings (see Table 6). Accordingly, we concluded that our measurement model's discriminant validity was satisfactory.

Table 5: Fornell-Larcker Criterion.

| Factor | PEWG | PIWG |
| --- | --- | --- |
| PEWG | **0.929** | |
| PIWG | 0.587 | **0.931** |

Note: Diagonal entries in bold type are the square root of the AVE.

Table 6: Cross Loadings.

| Item | PEWG | PIWG |
| --- | --- | --- |
| PEWG1 | 0.955 | 0.559 |
| PEWG2 | 0.900 | 0.544 |
| PEWG3 | 0.932 | 0.536 |
| PIWG1 | 0.568 | 0.935 |
| PIWG2 | 0.527 | 0.913 |
| PIWG3 | 0.543 | 0.945 |

### 3.2 Concurrent Validity and Sensitivity

Given that our model had acceptable convergent validity as well as suitable reliability and discriminant validity, we felt confident applying the manifest variables to investigate the concurrent validity and sensitivity of our warm-glow constructs. A Pearson product-moment correlation was used in order to determine whether exposure to EWG and IWG, which we simulated through the use of vignettes, impacts the values of PEWG and PIWG. The results revealed that moderate concurrent validity existed for PIWG with respect to the presence of IWG, $r(340) = 0.395$, $p < 0.001$. For PEWG, we established that weak concurrent validity exists with respect to the presence of IWG, $r(340) = 0.216$, $p < 0.001$. We found no statistically significant validity for either PEWG or PIWG with respect to EWG. In other words, PEWG and PIWG were successfully able to capture the effect we simulated through the use of the IWG vignette, but not the EWG vignette. The results of the analysis are visible in Table 7.

Table 7: Correlations.

| Factor | EWG | IWG | PEWG | PIWG |
| --- | --- | --- | --- | --- |
| EWG | 1.000 | 0.011 | 0.029 | 0.011 |
| IWG | 0.011 | 1.000 | 0.216* | 0.395* |
| PEWG | 0.029 | 0.216* | 1.000 | 0.587* |
| PIWG | 0.011 | 0.395* | 0.587* | 1.000 |

*$p < 0.01$.

In order to ascertain the sensitivity of the PEWG and PIWG constructs when exposed to EWG and IWG, we applied the approach prescribed by Rosengren et al. [28]. Specifically, we used a Multivariate Analysis of Variance (MANOVA) to examine whether each vignette had significantly affected how participants responded to questions. We first investigated whether Tabachnick and Fidell's [30] assumptions with respect to the MANOVA's were satisfied. We found that five of the assumptions were indeed satisfied. First, we confirmed that the dependent variables, PEWG and PIWG, were continuous. Second, we confirmed that the independent variables formed two categorically independent groups: EWG and IWG. Certainly, both were at two levels (exposed or not exposed). Third, we confirmed that our observations were independent, as participants were included only in one group. Fourth, we found that the dependent variables were positively correlated, $r(340) = 0.587$, $p < 0.001$, satisfying the conditions for linearity and multicollinearity [30–32]. Finally, in keeping with DeCarlo's [33] implementation of the multivariate test for SPSS, we tested the assumptions of multivariate outliers. The results revealed that there were no multivariate outliers, as the largest Mahalanobis distance (sq = 15.87) was lower than the critical values (Bonferroni) for a single multivariate outlier (critical $F(.05/n) = 17.25$, df = 2, 337).

However, the two remaining assumptions were not satisfied. Specifically, the equality of covariances assumption—tested using Box's test ($p = 0.025 < 0.05$)—found that the assumption was not satisfied. Conversely, again using DeCarlo's [33] implementation of the multivariate test for SPSS, we tested the assumptions of multivariate normality. Mardia's test for multivariate normality ($p = 0.026$) was found to have been violated, which introduced a concern with respect to the stability of the results. To address these violations we considered the words of Konietschke et al. [34], who recommend the use of bootstrap as a workaround. Specifically, they write "for moderate sample sizes, the bootstrap approach provides an improvement to existing methods in particular for situations with nonnormal data and heterogeneous covariance matrices in unbalanced designs". Moreover, as per Davison and Hinkley [35], the use of the bootstrapping technique ensures the stability of our results.

The bootstrapped MANOVA (2000 iterations) results revealed that when we selected IWG as a fixed variable, both the Wilks' Lambda (Wilks' Lambda = 0.842, $F(2, 335) = 31.450$, $p < 0.001$), which is a more traditional measure, and the Pillai's Trace tests (Pillai's Trace = 0.158, $F(2, 335) = 31.450$, $p < 0.001$), which is a more conservative measure, were statistically significant (see Table 8). We rely on Pillai's Trace tests primarily because of the advice of Finch [36], who writes, "across the four test statistics examined in the study, it was found that Pillai's trace was the most robust to violations of assumptions". Thus, we concluded that at least one of the ratings for PIWG and PEWG had statistically significant differences, once we exposed participants to different levels of IWG. When we selected EWG as a fixed variable, neither the Wilks' Lambda nor Pillai's Trace tests were found to be statistically significant. In other words, neither PIWG nor PEWG ratings yielded statistically significant differences when we exposed participants to different levels of EWG. Furthermore, when EWG and IWG interacted, there were statistically significant interaction effects for both the PIWG and PEWG ratings. The MANOVA results revealed both the Wilks' Lambda (Wilks' Lambda = 0.981, $F(2, 335) = 3.166$, $p = 0.043$) and the Pillai's Trace tests (Pillai's Trace = 0.019, $F(2, 335) = 3.166$, $p = 0.043$) were statistically significant (see Table 8).

Table 8: Multivariate Tests.

| Source | Test | Value | F | H. d.f. | Error d.f. | Sig. |
|---|---|---|---|---|---|---|
| EWG | Wilks' Lambda | 0.999 | 0.158 | 2.000 | 335.000 | 0.854 |
| | Pillai's Trace | 0.001 | 0.158 | 2.000 | 335.000 | 0.854 |
| IWG | Wilks' Lambda | 0.842 | 31.450 | 2.000 | 335.000 | < 0.001 |
| | Pillai's Trace | 0.158 | 31.450 | 2.000 | 335.000 | < 0.001 |
| EWG * IWG | Wilks' Lambda | 0.981 | 3.166 | 2.000 | 335.000 | 0.043 |
| | Pillai's Trace | 0.019 | 3.166 | 2.000 | 335.000 | 0.043 |

Since the MANOVA revealed statistical significance for the independent variable IWG as well as a statistically significant interaction effect between the two independent variables (i.e., EWG and IWG) on the combined dependent variables (i.e., PEWG and PIWG), the next step involved carrying out a univariate Analysis of Variance (ANOVA) with respect to the dependent variables. However, we should draw attention to the fact that two of the assumptions required for one to execute an ANOVA were violated, and therefore will be addressed in greater detail below, even though according to Blanca et al. [37] an ANOVA is said to be robust with respect to violations of its assumptions.

The first violation concerned the heterogeneity assumption, examined through the use of Levene's Test, which was not satisfied for PIWG (Levene's statistic = 3.028, $p = 0.030 < 0.05$), indicating that the variances were not equal. We should note that for PEWG (Levene's statistic = 2.313, $p = 0.076 > 0.05$) the assumption was satisfied, indicating that the variances were equal. For guidance, we looked to the work of Blanca et al. [38] who point out that when dealing with groups of a similar size, violations of the assumption cease to be a concern. Expressly, the authors write, "our findings are consistent with the early research suggesting that balanced designs can be used as protection against the effect of variance heterogeneity" [38].

The second violation concerned the normality assumption. A Kolmogorov-Smirnov test indicated that PEWG did not follow a normal distribution, $D(340) = 0.124$, $p < 0.001$. Similarly, a Kolmogorov-Smirnov test also indicated that PIWG did not follow a normal distribution, $D(340) = 0.121$, $p < 0.001$. For guidance, we looked to Hong et al. [39] who write "it is known that ANOVA yields robust and accurate p-values, even when the normality assumption is violated", referencing Blanca et al. [37] and Schmider et al. [40]. In particular, Schmider et al. [40] write, that "the commonly given advice to use samples of 25 participants per condition in ANOVA designs to circumvent possible negative influences of violations of normality assumptions seems well heeded". Given that we have a greater number of participants in each of our groups, well above the stated minimum of 25, we feel confident to proceed.

The results of the univariate ANOVA—with respect to IWG for each PEWG $F(1,335) = 17.106$, $p < 0.001$; and IWG for each PIWG $F(1, 335) = 62.991$, $p < 0.001$—were statistically significant at the adjusted 0.025 (= 0.05/2) alpha level using the Bonferroni procedure to protect against Type I error inflation, following the recommendation of Armstrong [41]. Similarly, there was a statistically significant interaction between the effects of EWG and IWG on PEWG, $F(2, 335) = 15.694$, $p = 0.014$, again using the Bonferroni procedure to protect against Type I error inflation. These values can be seen in Table 9.

Table 9: Test of Between-Subject Effects.

| Source | Dependent Variable | Type III Sum of Squares | d.f. | Mean Square | F | Sig. |
| --- | --- | --- | --- | --- | --- | --- |
| IWG | PEWG | 44.094 | 1 | 44.094 | 17.106 | < 0.001 |
| | PIWG | 119.530 | 1 | 119.530 | 62.991 | < 0.001 |
| EWG * IWG | PEWG | 15.694 | 1 | 15.694 | 6.088 | 0.014 |
| | PIWG | 6.105 | 1 | 6.105 | 3.217 | 0.074 |

We can conclude that both the PIWG and PEWG ratings were different between participants who were exposed to IWG, and those who were not. Specifically, the presence of IWG led to an increase in the averages of PIWG (from 4.645 to 5.827) and PEWG (from 3.632 to 4.347), compared to the control group. The presence of EWG did change the value of PEWG (from 3.948 to 4.042) and PIWG (from 5.223 to 5.262), but those changes were not statistically significant. Furthermore, a statistically significant interaction effect appeared to exist between EWG and IWG with respect to PEWG. Specifically, the presence of IWG without EWG led to the highest perception of PEWG (4.520). The presence of both EWG and IWG led to a decrease in that perception of PEWG (from 4.520 when exposed solely to IWG to 4.184 when exposed to both EWG and IWG). The descriptive statistics can be seen in Table 10.

Table 10: Descriptive Statistics.

| | EWG | IWG | N | Mean | Std. Dev. |
| --- | --- | --- | --- | --- | --- |
| PEWG | 0 | 0 | 83 | 3.369 | 1.536 |
|       |       | 1     | 84  | 4.520 | 1.459 |
|       |       | Total | 167 | 3.948 | 1.601 |
|       | 1     | 0     | 84  | 3.893 | 1.671 |
|       |       | 1     | 89  | 4.184 | 1.732 |
|       |       | Total | 173 | 4.042 | 1.704 |
|       | Total | 0     | 167 | 3.633 | 1.622 |
|       |       | 1     | 173 | 4.347 | 1.610 |
|       |       | Total | 340 | 3.996 | 1.653 |
|       | 0     | 0     | 83  | 4.498 | 1.435 |
|       |       | 1     | 84  | 5.952 | 1.071 |
|       |       | Total | 167 | 5.230 | 1.457 |
| PIWG  | 1     | 0     | 84  | 4.790 | 1.567 |
|       |       | 1     | 89  | 5.708 | 1.390 |
|       |       | Total | 173 | 5.262 | 1.544 |
|       | Total | 0     | 167 | 4.645 | 1.505 |
|       |       | 1     | 173 | 5.827 | 1.248 |
|       |       | Total | 340 | 5.246 | 1.500 |

## 4 DISCUSSION AND CONCLUSIONS

Through this work we offer two constructs that can be used within the practice of technology adoption modeling, and specifically the 'standard' models (i.e., TAM3 and UTATU2). They are designed specifically for understanding technology products that evoke a feeling of warm-glow in the consumer, which we referred to as 'good tech'. It has been found that "people's motivation for their giving behaviors" is either "impurely altruistic—motivated by the prospect of awards such as praise and respect—or purely altruistic—motivated by a genuine joy of giving" [42]. Accordingly, we adapted two constructs to measure individuals' perceptions of the two dimensions of warm-glow (i.e., extrinsic and intrinsic), PEWG and PIWG. Both constructs are rooted within the existing literature. The individual items for the PEWG construct were based on Han, Lee and Hwang's [28] factor of "social value, status", while the individual items for the PIWG construct were adapted from Müller, Fries and Gedenk [27].

The analysis revealed that both constructs do indeed capture their respective dimensions of warm-glow (i.e., extrinsic through the use of PEWG and intrinsic through the use of PIWG). However, our attempts to evoke EWG explicitly did not lead to a statistically significant change in participants' perception of EWG, only to a minor positive change. This leads us to conclude that even though end-users appear to be attracted to products that nourish their ego, they want this to be subtle, suggesting that they do not derive pleasure from engaging in activities that overtly try to target and satisfy their vanity. Two possibilities might serve as explanations of what we observed. First, the extrinsic reward that the vignette presents—recognition through social media—is not one that participants regard as valuable. Second, the participants value their privacy and do not want any data to be shared through social media, even if it might be considered a positive activity. This finding is not extensively described in the literature, but relates to the work of Winterich [43]. This author organized annual group donations in memory of a family member, only to find that most individuals in her circle who made donations for a common cause did not want their contributions publicly announced through social media (specifically, Facebook).

Furthermore, attempts to induce IWG explicitly led to the presence of EWG, leading us to the question of whether one form of warm-glow can manifest itself inside a consumer without the presence of the other. This finding indicates that the relationship between EWG and IWG is very close and the two can be considered as two sides of the same coin. At some level, possibly even subconsciously, there is a selfish component associated with doing good inside the user's mind. Individuals likely formulate ways in which their ego can benefit, even when they outwardly present their motivations as non-altruistic. This mirrors what has been reported in the literature. Evren and Minardi [44] write that "an individual who experiences warm-glow values the availability of selfish options even if she plans to act unselfishly". The Decision Lab [42] brings attention to Andreoni [45] by writing "to this end, everyone who experiences warm glow—assuming the model is accurate—is self-deceiving". They then go on to write, "however, the criticism should recognize that according to the original model, in order for warm glow to exist, pure altruism is not possible, as it is inherently self-motivated" [42].

## 4.1 Implications

This study contributes to the general technology adoption literature by offering constructs to be used in the modeling of the adoption of 'good tech'. We see an increase in technology products that could potentially fall into that category and would evoke such a feeling in consumers, albeit there being no one universal perception of 'goodness'—this, in essence, depends on an individual's values. Indeed, what may appear good to one individual may not seem that way to another. For example, a user who has an interest in fighting AIDS would conceivably perceive Apple's iPhone Red as 'good', whereas a user who has an interest in planting trees would most likely perceive the web-based search tool Ecosia as 'good', and a user who has an interest in fighting global warming would probably perceive Toyota's Prius line of automobiles as 'good'. As the warm-glow phenomenon has the ability to lead to an "enhanced willingness to buy", and that the "effects persist and play out in actual behavior" [46], it can influence consumer decisions and lead to novel consumption patterns [47]. Correspondingly, being able to measure its perceived presence is of value, particularly for those involved in offering, or considering offering, products that evoke a feeling of warm-glow in consumers.

## 4.2 Limitations and Future Research Directions

There are two limitations that we feel should be highlighted. The first has to do with our sample being comprised solely of individuals from the United States. Looking to the literature we find reports that the extent to which warm-glow impacts user decisions varies based on the culture of those consumers [48]. Consequently, future research may want to investigate the constructs' suitability for use in other cultures (i.e., outside of the United States). The second limitation concerns the characteristics of the technology we used to test our model for this study: a generic (unbranded) internet search solution, which is a standard, easy-to-use technology that is frequently offered without charge. Products with different characteristics regarding familiarity, complexity, price, and brand may lead participants to perceive warm-glow differently. Consequently, research investigating the interplay between these factors and the presence of warm-glow would provide valuable insight. Finally, we conclude by pointing out that that next step to develop this research further would be to integrate and evaluate both constructs for the two contemporary models—specifically, TAM3 and UTAUT2.

## ACKNOWLEDGMENTS

This research was funded in part by a New York University School of Professional Studies Dean's Research Grant.